\documentclass[a4paper]{aa}
\usepackage{graphicx}
\usepackage{txfonts}
\usepackage{amssymb}
\begin{document}

  \title{Large-scale molecular shocks in galaxies: the SiO interferometer map of \object{IC~342}\thanks{Based 
  on observations carried out with the IRAM Plateau de Bure Interferometer. IRAM
  is supported by INSU/CNRS (France), MPG (Germany) and IGN (Spain).}}

\authorrunning{A. Usero et al.}
\titlerunning{Large-scale molecular shocks in galaxies}

\author{A. Usero\inst{1}, S. Garc\'{\i}a-Burillo\inst{1}, J.
Mart\'{\i}n-Pintado\inst{2}, A. Fuente\inst{1} and R. Neri\inst{3}}

\offprints{A. Usero,\\
\email{a.usero@oan.es}}

\institute{Observatorio Astron\'omico Nacional (OAN) --Observatorio de Madrid--, C/ Alfonso XII 3,
  28014 Madrid, Spain\\
e-mail: {\tt a.usero@oan.es, s.gburillo@oan.es, a.fuente@oan.es}
 \and Instituto de Estructura de la  Materia, DAMIR-CSIC, C/ Serrano 121, 28006  Madrid, Spain\\
e-mail: {\tt jmartin.pintado@iem.cfmac.csic.es} 
\and IRAM, 300 rue de la Piscine, Domaine Universitaire, 38406 St. Martin d'H\`eres Cedex, France\\
e-mail: {\tt neri@iram.fr}
}

   \date{Received 12 August 2005; Accepted 6 October 2005}

\abstract{
We present the first high-resolution ($5\farcs6\times5\farcs1$) images of the emission of silicon monoxide
(SiO) in the nucleus of the nearby spiral \object{IC~342} obtained with the IRAM Plateau de Bure
Interferometer (PdBI). Using a two-field mosaic, we have mapped simultaneously the emission of the
SiO($v=0,~J=2-1$) and H$^{13}$CO$^+$($J=1-0$) lines  in a region of { $\sim$0.9~kpc~$\times$~1.3~kpc}
(RA$\times$Dec)--size centered around the nucleus of \object{IC~342}. The bulk of the emission in the two
lines comes from a { ${\sim290}$~pc} spiral arm located North and a central component that forms the southern
ridge of a { ${r\sim80}$~pc} nuclear ring which has been identified in other interferometer maps of the
galaxy. We detect continuum emission at 86.8~GHz in a { ${\sim80-180}$~pc} central source. The continuum
emission, dominated by thermal free-free bremsstrahlung, is mostly anticorrelated with the observed
distribution of SiO clouds.  The SiO--to--H$^{13}$CO$^+$  intensity ratio is seen to increase by an order of
magnitude from the nuclear ring ($\sim0.3$) to the spiral arm ($\sim3.3$). Furthermore the gas kinematics show
significant differences between SiO and H$^{13}$CO$^+$ over the spiral arm where the linewidths of SiO are a
factor of 2 larger than that of H$^{13}$CO$^+$. The average abundance of SiO in the inner { ${r\sim320}$~pc}
of \object{IC~342} is X(SiO)$\gtrsim2\times10^{-10}$. This evidences that shock chemistry is at work in the
inner molecular gas reservoir of \object{IC~342}.    

To shed light on the nature of shocks in \object{IC~342}, we have compared the emission of SiO with another
tracer of molecular shocks: the emission of methanol (CH$_3$OH).  We find that the significant difference of
the abundance of SiO measured between the spiral arm (X(SiO)$\sim$a few $10^{-9}$) and the nuclear ring
(X(SiO)$\sim$10$^{-10}$) is not echoed by a comparable variation in the SiO--to--CH$_3$OH intensity
ratio. This implies that the typical shock velocities should be similar   in the two regions. In contrast, the
fraction of shocked molecular gas should be $\sim5-7$ { times} larger in the spiral arm (up to $\sim$10$\%$ of
the   available molecular gas mass over the arm region) compared to the nuclear ring. In the light of these
results, we revise the validity of the various scenarios which have been proposed to explain the onset of
shock chemistry in galaxies and study their applicability to the nucleus of \object{IC~342}.   We conclude
that the large-scale shocks revealed by the SiO map of \object{IC~342} are mostly unrelated with star
formation and  arise instead in a pre-starburst phase. Shocks are driven by cloud-cloud collisions along the
potential well of the \object{IC~342} bar. The general implications for the current understanding of galaxy
evolution are discussed.  
} 

   \maketitle
%

\section{Introduction}
\label{secint}

There is mounting evidence that the properties of molecular gas in starbursts (SBs) and Active Galactic Nuclei
(AGNs) differ from that of quiescent star forming galaxies (e.g., Genzel et al.~\cite{genz98}). The
spectacular energies  injected into the gas reservoirs   of {\it active} galaxies can create a particularly
harsh environment for the neutral ISM. Although  thus far restricted to a handful of objects, multiline
millimeter wave studies have allowed to study the onset of large-scale shocks, the propagation of chemistry of
Photon Dominated Regions (PDRs) or the prevalence of X-ray Dominated Regions (XDRs) in the molecular disks of
SBs and AGNs (e.g., Mauersberger \& Henkel~\cite{maue93}; Tacconi et al.~\cite{tacc94}; Garc\'ia-Burillo et
al.~\cite{buri00,buri01b,buri02}; Mart\'{\i}n et al.~\cite{mart03,   mart05}; Usero et al.~\cite{user04};
Fuente et al.~\cite{fuen05a}; Meier \& Turner~\cite{meie05}).   

The first SiO($v=0$, $J=2-1$) maps made with the Plateau de Bure Interferometer (PdBI) in the nuclei of the
prototypical starbursts \object{NGC~253} and \object{M~82} have revealed the existence of large-scale
molecular shocks in galaxy disks (Garc\'ia-Burillo et al.~\cite{buri00,buri01b}). Different   scenarios have
been proposed to account for the emission of SiO in our own Galaxy and in the nuclei of external  galaxies. On
small $\sim$pc--scales, studies in the Galaxy disk show that the enhancement of SiO in gas phase can be
produced in the bipolar outflows of young stellar objects (YSOs), due to the sputtering of dust grains by
shocks (Mart\'{\i}n-Pintado et al.~\cite{mart92}; Schilke et al.~\cite{schi97}; Cesaroni et
al.~\cite{cesa99}). On larger scales, Mart\'{\i}n-Pintado et al. (\cite{mart97}) reported the detection of  a
SiO $\sim$150~pc circumnuclear disk (CND) in the Galactic Center region. In this CND high fractional
abundances of SiO are found in molecular clouds which are not actively forming stars, but where bar models for
our Galaxy predict a high likelihood for cloud collisions (H\"uttemeister et al. \cite{huet98}). In
\object{M~82}, virtually all of the SiO emission traces the disk-halo interface where episodes of mass
injection are building up the gaseous halo (Garc\'ia-Burillo et al.~\cite{buri01b}). Garc\'ia-Burillo et
al.~(\cite{buri00}) have discussed the role of bar resonances at inducing shocks in  the $\sim$600~pc CND of
\object{NGC~253}. However, the high-inclination of \object{NGC~253} and the limited perspective of the
Galactic Center region make the determination of bar resonance positions mostly dependent on kinematical models.  

In this paper we study at high-resolution ($\sim$5\arcsec) the SiO($v=0,~J=2-1$) emission in the inner
$r\sim30\arcsec$ of \object{IC~342}, using the IRAM Plateau de Bure Interferometer (PdBI). \object{IC~342} is
a nearby { (D$\simeq$3.3~Mpc, i.e., 1\arcsec$\simeq$16~pc; Saha, Claver \& Hoessel \cite{saha02})} weakly
barred spiral galaxy which hosts a moderate starburst episode in the central { ${r\sim80}$~pc} nuclear region
(B\"oker et al. \cite{boke97, boke99}). Thanks to its nearly face-on orientation and close distance,
\object{IC~342} is an   optimal testbed where the underlying mechanisms of large-scale molecular shocks can be
probed using the PdBI. At this distance, the PdBI can trace and spatially resolve the SiO emission of shocked
molecular gas on scales of individual GMCs { (${\sim80}$~pc)} in the disk of \object{IC~342}. Several works
have underlined the similarities between the nucleus of our Galaxy and that of \object{IC~342} in terms of the
measured gas mass fractions, stellar masses and star formation (Downes et al.~\cite{down92}). This study can
thus help to shed light on the  origin of molecular shocks in the nucleus of our own Galaxy.

The distribution of molecular gas in the inner { ${r\sim320}$~pc} of \object{IC~342}, revealed by the
published interferometer CO maps of the galaxy, is reminiscent of the typical response of gas to a bar
potential (Lo et al. \cite{lo84}; Ishizuki et al. \cite{ishi90};  Levine et al. \cite{levi94}; Meier \& Turner
\cite{meie01, meie05}; Schinnerer et al.~\cite{schi03}). Two gas lanes are shifted with respect to the 
major axis of the { ${\sim9}$~kpc} bar oriented with a position angle $\sim20\degr$ (Buta \&
McCall~\cite{buta99}).  The gas lanes delineate a two-arm {\it spiral} pattern that ends at a nuclear {
  ${r\sim80}$~pc} {\it ring}. This pattern would correspond to the transition from x$_1$ orbits (outer disk)
down to x$_2$ orbits (inner disk) of the \object{IC~342} bar, assuming that an Inner Lindbland Resonance (ILR)
exists near { ${r\sim80-160}$~pc.}  

Previous interferometer maps have probed the dense gas content of the nucleus of \object{IC~342} (Ho et
al.~\cite{ho90}: NH$_3$; Downes et al.~\cite{down92}: HCN; Nguyen-Q-Rieu et  al.~\cite{nguy92}: HCO$^+$; Meier
\& Turner~\cite{meie05}: HNC, HC$_3$N, C$_2$H, C$^{34}$S, HNCO, CH$_3$OH and N$_2$H$^+$). Five major GMCs,
labelled originally as A--to--E in the HCN map of Downes et al. (\cite{down92}) and later redefined by Meier
\& Turner~(\cite{meie01}), are identified in the inner { ${r\sim320}$~pc disk}. Meier \&  Turner (\cite{meie05})
find remarkable differences in morphology between  the various molecular emission maps of
\object{IC~342}. This is interpreted as an evidence of strong chemical differentiation in the nucleus of
\object{IC~342}. While some molecules trace the Photon Dominated Regions close to the nuclear starburst
{ (${r\sim80-180}$~pc)}, the emission of molecular species such as methanol (CH$_3$OH) is considered to be
stemming from shocks (Meier \&  Turner~\cite{meie05}).  

In the scenario of shock chemistry there is evidence that CH$_3$OH and SiO trace distinctly different velocity
regimes in shocks. SiO is seen to be associated with more energetic events, i.e., those potentially more
efficient at processing dust grains (Garay et al. \cite{gara00}). In this paper we take advantage of the
complementarity  of SiO and CH$_3$OH as tracers of shock chemistry and use the SiO--to--CH$_3$OH ratio in
\object{IC~342} to discuss the origin of large-scale shocks in this galaxy. With this aim we make a
quantitative comparison of our results with those obtained in other well-known references for shock chemistry
in our Galaxy and in external galaxies.   

We describe in { Sect.~\ref{secobs}} the PdBI observations used in this paper. { Sect.~\ref{secres}} presents the main
results issued from the analysis of the continuum image and the SiO and H$^{13}$CO$^+$(1--0) line maps of
\object{IC~342}.   We derive in { Sect.~\ref{secabu}} the fractional abundances of SiO in the disk of the galaxy
and interpret their relation with the published CH$_3$OH map in { Sect.~\ref{seceff}}. { Sect.~\ref{secori}} discusses
the possible mechanisms driving large-scale shocks in \object{IC~342}. In { Sect.~\ref{secwav}} we analyze the
potential role of density waves as drivers of the large-scale shock chemistry in \object{IC~342}.  The main
conclusions are summarized in { Sect.~\ref{seccon}}.


\section{Observations}
\label{secobs}

Observations of \object{IC~342} were carried out with the PdBI from July to August 2001. We observed
simultaneously the J=2--1 line of SiO  (86.847~GHz) and the J=1--0 line of H$^{13}$CO$^+$ (86.754~GHz) using
the CD set of configurations. The primary beam of the PdBI at 87~GHz is 55\arcsec.  Two positions shifted
$(0\arcsec,-12\arcsec)$ and $(0\arcsec,+12\arcsec)$ from the phase center,
$\alpha_\mathrm{J2000.0}$=03$^{\mathrm{h}}$46$^{\mathrm{m}}$48\fs01,
$\delta_\mathrm{J2000.0}$=68\degr05\arcmin46\farcs0, were observed in mosaic mode.   We adjusted the spectral
correlator to give a contiguous bandwidth of 1500~km~s$^{-1}$. The frequency resolution was set to  1.25~MHz
(4.3~km~s$^{-1}$) during the observations; channels were resampled to a velocity resolution of 5~km~s$^{-1}$
in the final maps. We calibrated visibilities using \object{0224+671} as amplitude and phase reference. The
absolute flux scale was derived on \object{MWC~349}, and receiver passband was calibrated on \object{3C~454.3}
and \object{3C~345}. 

Mosaics were CLEANed using the MAPPING procedure of the GILDAS software  package, which includes primary beam  
correction. The synthesized clean beam is 5\farcs6$\times$5\farcs1 size (PA=321\degr) for the line
maps. Images are 300~$\times$~300 pixels in extent, with a pixel size of 0\farcs35.   The rms noise level in
5~km~s$^{-1}$ wide channels,  derived after subtraction of the continuum emission, is 1.5~mJy~beam$^{-1}$ at
the center of the maps.

A 3.5~mm continuum map was generated averaging channels free of line emission.
Uniform weighting was applied to the measured visibilities, producing a  clean beam of
$4\farcs1\times3\farcs8$  (PA~=~$111\degr$). The rms at the center is 0.45~mJy~beam$^{-1}$.


\section{Results}
\label{secres}

\subsection{The 3.5~mm continuum map}
\label{subcon} 

\begin{figure}
\includegraphics[angle=-90,width=\hsize]{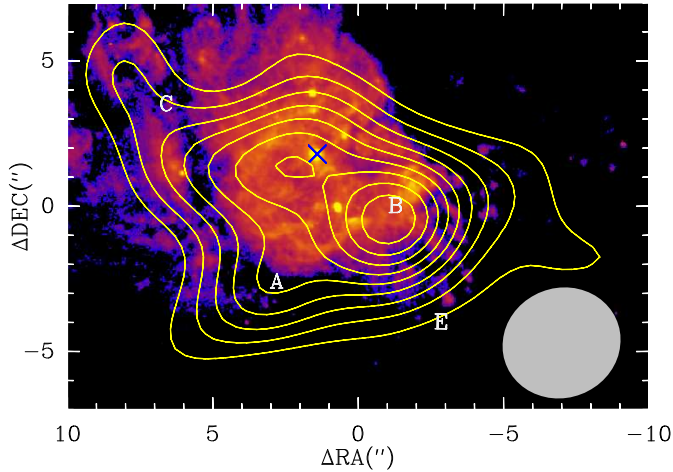}
\caption{
We overlay the 3.5~mm continuum map obtained with the PdBI (contour levels: 0.90 to 4.5 in steps of
$0.45$~mJy~beam$^{-1}$) with the H$\alpha$+continuum image (color scale in arbitrary units) obtained by HST
towards the nucleus of \object{IC~342}. Positions of GMCs A, B, C and E (as defined by Meier \&
Turner~\cite{meie01}), are highlighted; the position of GMC D, beyond the displayed region of this figure, has been
redefined in this work according to the local maximum of SiO integrated intensity (see Fig.~\ref{figint}). A
 cross marks the position of the  
galactic center determined by Schinnerer et al. (\cite{schi03}).  Relative positions are referred to the phase
center $\mathrm{\alpha_{J2000.0}=03^h46^m48^s.01}$,  $\delta_{J2000.0} = 68\degr05\arcmin46\farcs0$.  The
$4\farcs1\times3\farcs8$ beam is drawn at the bottom--right corner.   
}     
\label{figcon}
\end{figure}

\begin{figure*}
\includegraphics[angle=-90,width=\hsize]{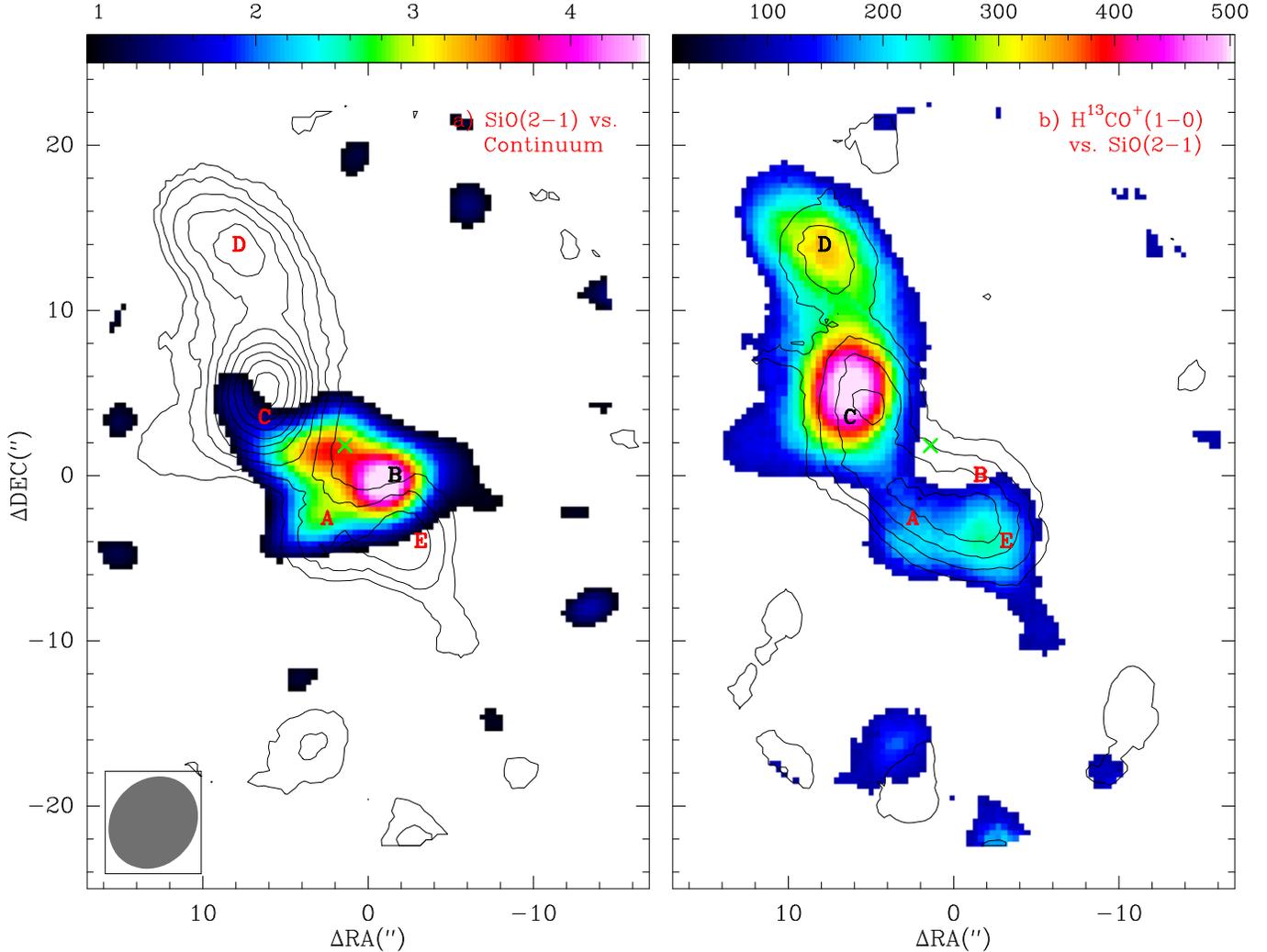}
\caption{
{\bf a(left)} We overlay the SiO(2--1) integrated intensity map (contour levels from 90 to 522 in steps of
54~mJy~beam$^{-1}$~km~s$^{-1}$; $1\sigma=36$~mJy~beam$^{-1}$~km~s$^{-1}$) with the 3.5~mm continuum map (color
scale from 0.90 to 4.5~mJy~beam$^{-1}$; $1\sigma=0.45$~mJy~beam$^{-1}$). {\bf
  b(right)} Same as {\bf a(left)} with H$^{13}$CO$^+$(1--0) (contours) overlaid with SiO(2--1) (color scale), both
displayed with the same intensity spacing (used in {\bf a(left)}). The two integrated emission maps were obtained
with a $1.2\sigma$ clipping applied to the velocity interval [$-10$, $85$]~km~s$^{-1}$. Line contours are
screened beyond the displayed field-of-view as S/N ratio is lower than 2.5 farther out. Beam-size is
represented by a filled ellipse. 
} 
\label{figint} 
\end{figure*}

The 3.5~mm continuum emission contours are displayed in Fig.~\ref{figcon}. The total flux integrated within
the inner { ${r\sim180}$~pc} of the galaxy is  $\sim$17~mJy. The flux recovered is of $\sim$22~mJy when 
derived from a lower 9$\arcsec$ resolution version of the PdBI map; the latter is obtained assuming a Gaussian
UV taper on the visibilities (with a FWHM=70~m). These values are roughly in agreement with the previous
estimates obtained by Downes et al.~(\cite{down92}) and  Meier \& Turner (\cite{meie01}) at similar
frequencies. The bulk of the continuum emission comes from the inner { ${r\sim80}$~pc} of \object{IC~342} and it is
closely linked to the nuclear star forming region identified in the H$\alpha$+continuum HST image of the galaxy
(Fig.~\ref{figcon}). As illustrated in Fig.~\ref{figint}, the continuum emission is mostly anticorrelated with
the emission coming from the dense molecular gas traced by SiO or H$^{13}$CO$^+$.

Despite its compactness, the continuum source is spatially resolved by the PdBI beam. The morphology of the
emission follows closely the distribution of star forming complexes in the nucleus of \object{IC~342}: two
5~Myr-old H\textsc{ii} complexes (B\"oker et al. \cite{boke97}) which are close to GMCs B and C 
(notation of Meier \& Turner \cite{meie01}), and an older (6-60~Myr) star cluster close to the center
of the galaxy (B\"oker et al. \cite{boke97,boke99}).  As shown in Fig.~\ref{figcon}, two peaks of emission shape the
morphology of an elongated disk oriented along PA$\simeq64\degr$. The principal peak is related to the western
H\textsc{ii} complex, which is identified in the HST map to be close to GMC B. The secondary maximum is close
to the dynamical center where the old star cluster is detected. Lower-level emission extends East from the
central disk towards GMCs A and C. The latter extension is close to the eastern  H\textsc{ii} complex. 

Accounting for the differences in spatial resolution and sensitivity, the morphology of the 3.5~mm continuum
PdBI map agrees with that of previous radiocontinuum maps obtained at other wavelengths (Condon et
al. \cite{cond82}: 21~cm; Turner \& Ho \cite{turn83}: 2 and 6~cm; Ho et al. \cite{ho90}: 1.3~cm; Downes et 
al. \cite{down92}: 3.4~mm). At higher frequencies, however, the contribution from dust to  the thermal emission may
not be negligible. This would explain the  differences between the continuum map of Fig.~\ref{figcon} and the
1.3~mm  continuum image obtained by Meier \& Turner (\cite{meie01}).

Based on measurements at various frequencies, previous works have concluded that thermal free-free
bremsstrahlung should dominate the emission budget at 3.5~mm (Downes et al. \cite{down92}; Turner \& Ho
\cite{turn83}; Turner \& Hurt \cite{turn92}). The good spatial coincidence between the 3.5~mm and the
H$\alpha$ emissions  supports the conclusion that the 3.5~mm emission traces the location of ongoing star
formation in the 
inner \textbf{${r\sim80}$~pc} of \object{IC~342}.       


\subsection{The line maps: SiO and H$^{13}$CO$^+$}

\subsubsection{Integrated intensity maps and line ratios}
\label{ratios}

Fig.~\ref{figint} shows the velocity-integrated intensity maps of SiO(2--1) and H$^{13}$CO$^+$(1--0) in  the
inner { ${r\sim320}$~pc} of \object{IC~342}. The spatial resolution of the PdBI map allows to resolve
the overall emission which, for both  tracers, is elongated along the N-S direction. Though with significant
differences between SiO and H$^{13}$CO$^+$, the basic  morphology of the maps is roughly in agreement with
that emerging from the HCN map of Downes et al. (\cite{down92}). In the case of SiO, we detect strong
emission in the spiral arm located North (GMCs C and D in Fig.~\ref{figint}). Weaker SiO emission
delineates the southern ridge of the nuclear ring (GMCs A and E in Fig.~\ref{figint}). Finally, SiO emission is not 
detected over the southern spiral arm. This result agrees with the overall picture derived from other dense
gas tracers in \object{IC~342} which are hardly detected over the southern spiral. Compared to SiO, 
the emission of H$^{13}$CO$^+$ is weaker along the northern spiral arm, while the peak of emission is found
on the southern ridge of the  nuclear ring.

\begin{figure}
\includegraphics[width=\hsize]{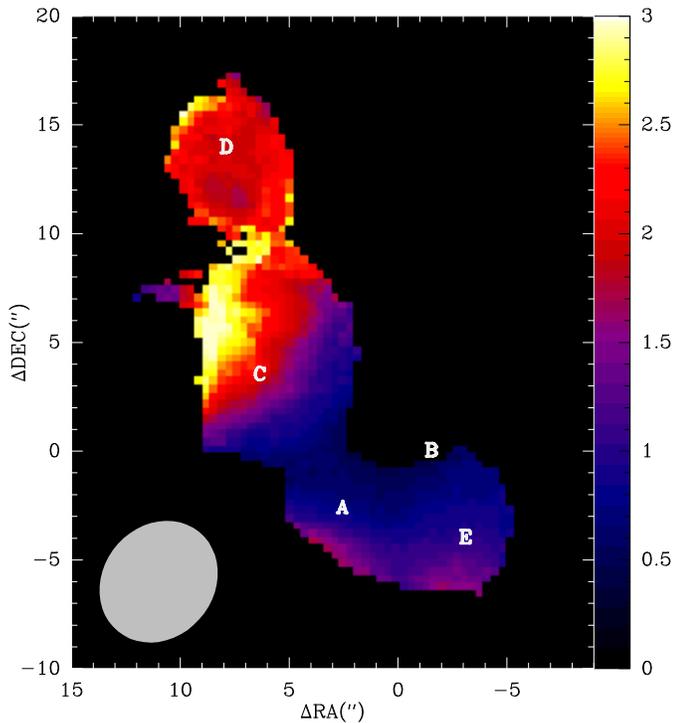}
\caption{
SiO(2--1)--to--H$^{13}$CO$^+$(1--0) integrated intensity ratio in the inner { ${r\sim320}$~pc} disk of
\object{IC~342}. The ratio is derived from the two maps of Fig.~\ref{figint}, assuming a 2.5$\sigma$-clipping
on the integrated intensities of both lines. The measured ratios (color scale) range from 0.3 to 3.3.
} 
\label{figrat}
\end{figure}

As it is shown in Fig.~\ref{figrat}, the different distributions of SiO and H$^{13}$CO$^+$ in \object{IC~342}
translate into an order of magnitude  difference in the SiO--to--H$^{13}$CO$^+$ intensity ratio, $R_I$, which
goes from $\sim$ 3.3 on the northern spiral arm down to $\sim$0.3 on the nuclear  ring. The average value of
$R_I$ inside the image field of view is $\sim$1.6. If we assume that the emission of both lines is optically
thin, the value of $R_I$ provides an estimate of the beam-averaged fractional abundance of SiO relative to
H$^{13}$CO$^+$,  which is accurate within a factor of 4 (see Sect.~\ref{secabu} for discussion). The reference
studies of galactic clouds (Mart\'{\i}n-Pintado et al.~\cite{mart92}; Bachiller \&
P\'erez-Guti\'errez~\cite{bach97}; Fuente et al.~\cite{fuen05b}) and external galaxies (Garc\'{\i}a-Burillo et
al. \cite{buri00, buri01b}; Usero et al.~\cite{user04}) indicate that a value of $R_I>$0.1 is a strong
indication that shock chemistry is at work in molecular gas. The reported values of $R_I$ for \object{IC~342}
are similar to those found on similar spatial  scales in the circumnuclear disk of \object{NGC~253}
($\sim$1-3;  Garc\'{\i}a-Burillo et al. \cite{buri00}) and, also, in the chimney and the supershell of
\object{M~82} ($\sim$0.4-3.5; Garc\'{\i}a-Burillo et al.~\cite{buri01b}).

Of particular note, $R_I$ reaches the largest values along the northern spiral arm, i.e., in a region of the
disk of \object{IC~342} where the evidence of active star formation is scarce. 
Regardless of its origin, the order of magnitude variation of $R_I$ measured over the disk of
\object{IC~342} reveals that shocks are processing molecular gas with a highly changing {
  \emph{efficiency} (in terms of the total mass of grain material processed by shocks relative to the total
  gas mass; see  discussion in Sects.~\ref{secabu}  and \ref{seceff}).}

\subsubsection{Gas kinematics and line profiles}
\label{subkin}

\begin{figure*}[thp!]
\sidecaption
\includegraphics[width=17cm]{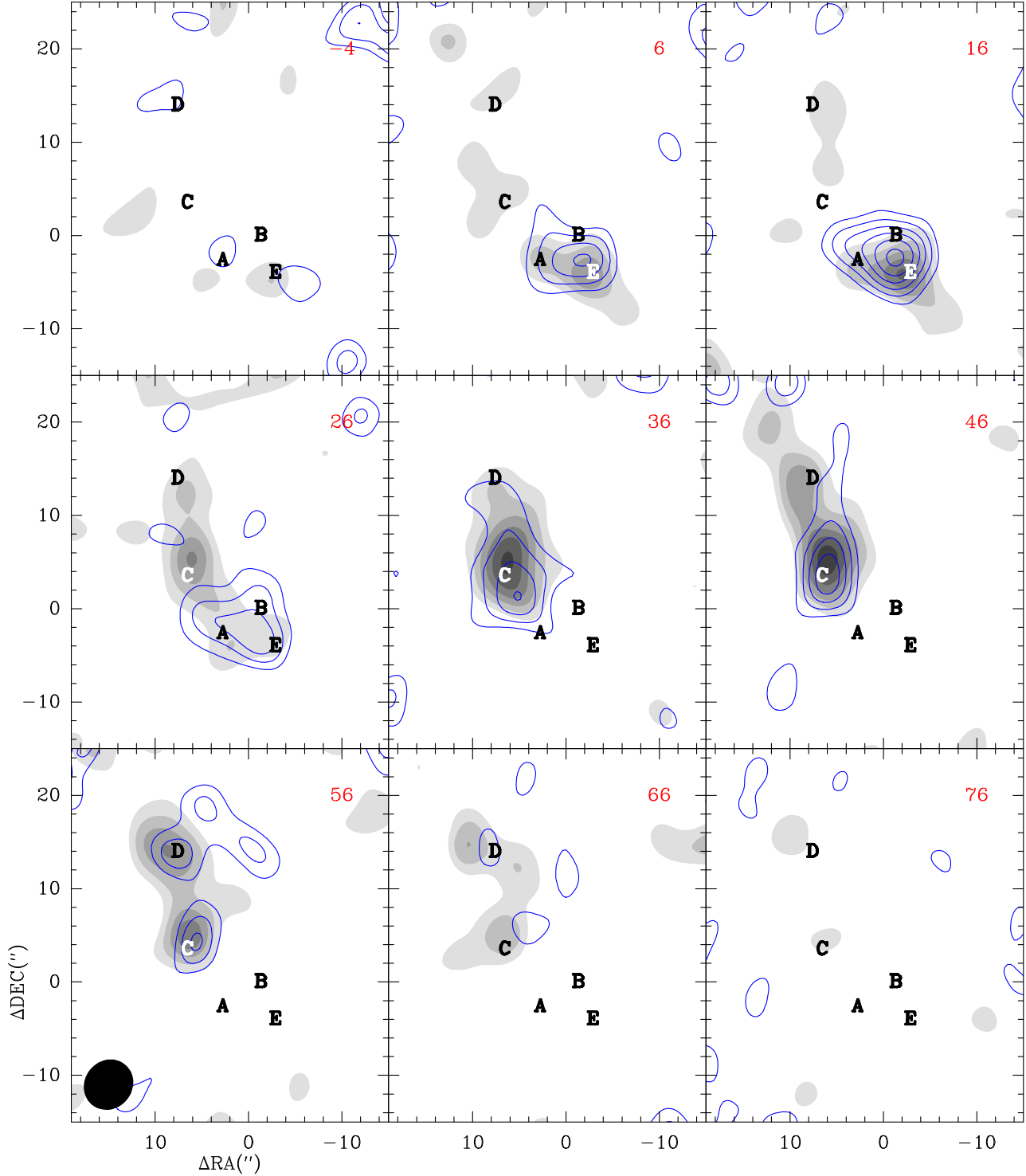}
\caption{
Velocity channel maps of SiO(2--1) (grey scale) and H$^{13}$CO$^+$(1--0) (contour levels) in \object{IC~342}. 
Both scales are from 2.5$\sigma$ to 10.5$\sigma$ in steps of 1.5$\sigma$;
$1\sigma=1.1$~mJy~beam$^{-1}$ in a 10~km~s$^{-1}$ channel. The channel
LSR-velocities and the positions of GMCs A--to--E are indicated in each panel.
}  
\label{figcha}
\end{figure*}

Fig.~\ref{figcha} shows the velocity-channel maps of SiO(2--1) and H$^{13}$CO$^+$(1--0) in the nucleus of
\object{IC~342}. For both tracers we see the expected velocity gradient due to the rotation of the
disk, which according to the fit of Crosthwaite et al.~(\cite{cros01}) should be maximal along the
kinematic major axis at PA=37$^{\circ}$.  The channel maps show that the gas kinematics are similar for SiO
and H$^{13}$CO$^+$ on the nuclear ring: the velocity centroids and  linewidths measured in SiO and
H$^{13}$CO$^+$ are roughly in agreement in this region (GMCs A, B, and E). However, the gas kinematics show
significant differences between the two species over the northern spiral arm (GMCs C and D):
the emission of SiO is detected from $v=16$~km~s$^{-1}$ to $v=66$~km~s$^{-1}$ near GMC D, i.e., roughly twice
the corresponding velocity interval for H$^{13}$CO$^+$. 

\begin{figure*}[htp]
\includegraphics[angle=0,width=\hsize]{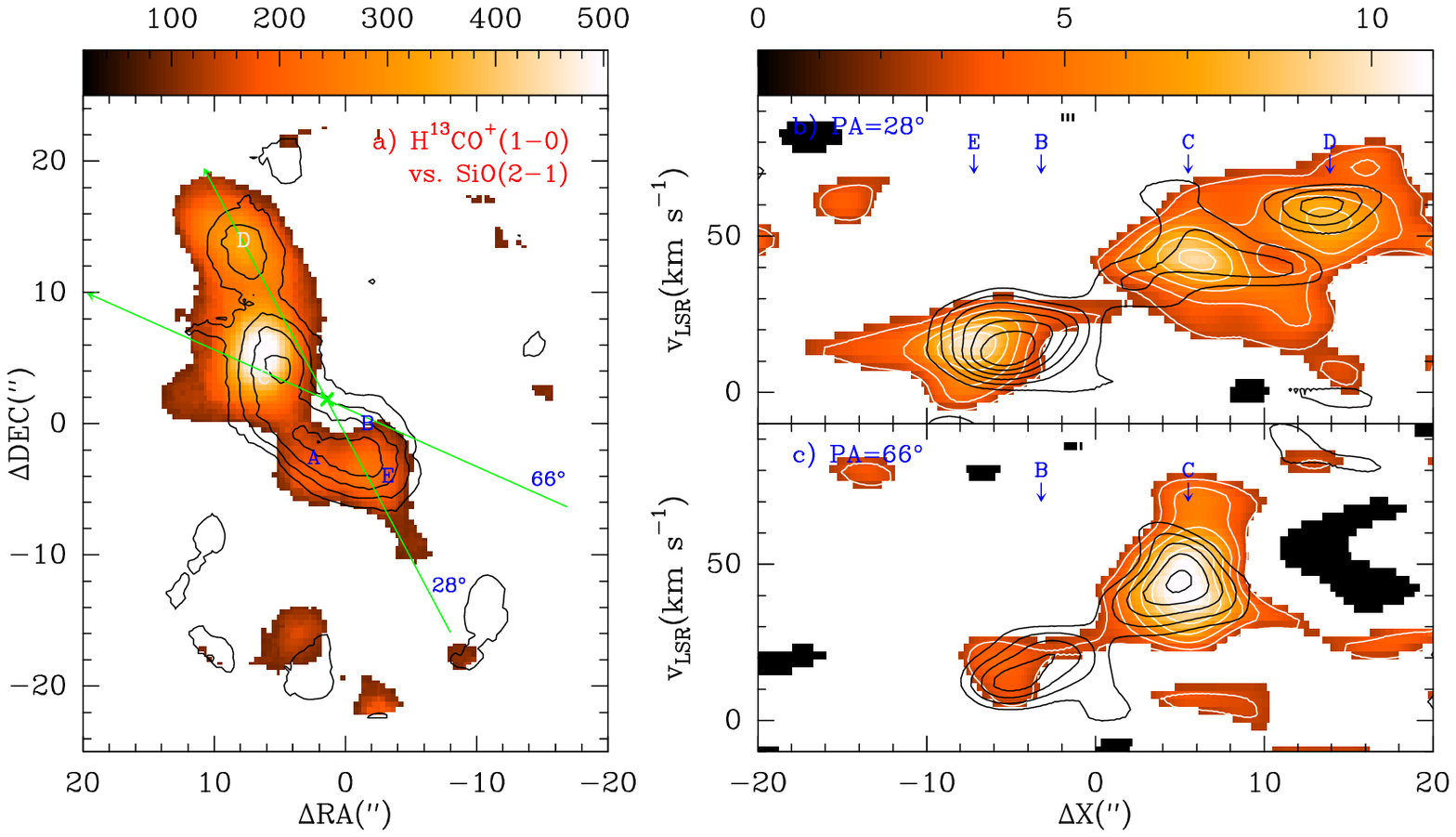}
\caption{
 SiO(2--1) and H$^{13}$CO$^+$(1--0) position-velocity (p--v) diagrams ({\bf b(top
  right)} and {\bf c(bottom right)}) taken along the {1D--strips} highlighted in {\bf a(left)} (overlay of the
SiO(2--1) and H$^{13}$CO$^+$(1--0) intensity maps as shown in Fig.~\ref{figint}b). { The position angles of
 the p--v diagrams are chosen to maximize the contrast between the linewidths measured in the northern
 spiral arm and in the nuclear ring.}
In the two p--v plots, the SiO(2--1) brightness is represented in color scale (2.6 to 11~mJy~beam$^{-1}$) and
 white contours (from 3 to 11 in steps of 1.5~mJy~beam$^{-1}$), while H$^{13}$CO$^+$(1--0) brightness levels
 appear in black contours (same levels as above). The approximate location of GMCs B,C, D and E along the
 strips is indicated in  {\bf   b(top right)} and {\bf c(bottom right)}. 
}  
\label{figsli}
\end{figure*}

The different kinematics of the SiO and H$^{13}$CO$^+$ lines over the northern spiral arm are illustrated by the
position--velocity plots shown in Fig.~\ref{figsli}. SiO lines become significantly wider than H$^{13}$CO$^+$
lines at the passage of the northern spiral arm GMCs (C and D), in contrast with the nuclear ring GMCs (B and
E) where linewidths are the same within the errors. While the velocity centroids derived in the two lines are
similar in the spiral arm region, the measured linewidths for  SiO are, on average, a factor of 2 larger than
the linewidths of H$^{13}$CO$^+$.  The estimated velocity dispersion of the GMCs ($\sigma_v$) is
$\sim20-25$~km~s$^{-1}$, in the northern spiral arm, and $\sim10$~km~s$^{-1}$, in the nuclear ring.  This
suggests that the  degree of {\it apparent} turbulence measured on GMC-like scales in the SiO emitting gas is
enhanced compared  to that of the more quiescent dense gas component traced by H$^{13}$CO$^+$ in this region.
The latter implies that the SiO(2--1)--to--H$^{13}$CO$^+$(1--0) ratio measured at the wings of the SiO lines 
should be even larger than the velocity-integrated ratio $R_I$ derived above for the northern spiral arm. This
would only reinforce the case of an enhanced SiO chemistry in gas phase in this region.

\begin{figure*}[btp]
\includegraphics[angle=-90,width=\hsize]{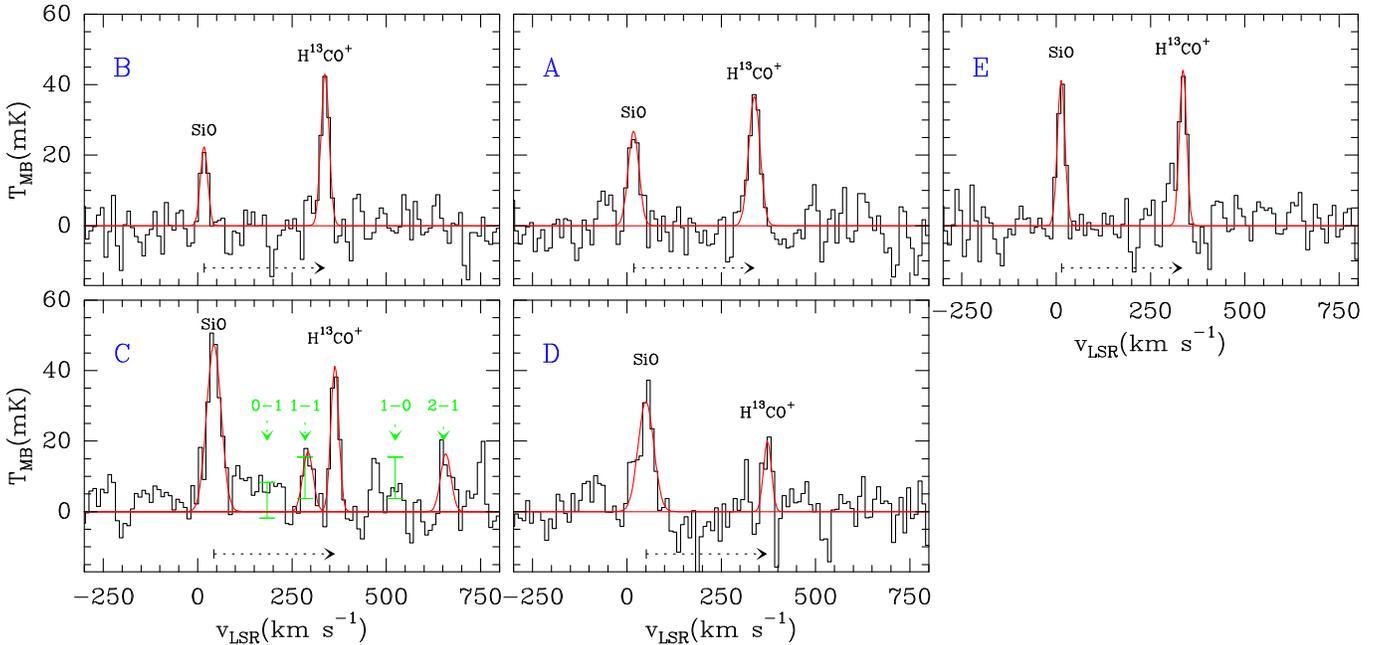}
\caption{
SiO(2--1) and H$^{13}$CO$^+$(1--0) spectra observed towards GMCs A--to--E. 
Spectra are smoothed to 10~km~s$^{-1}$ resolution. 
LSR-velocities are referred to the frequency  of SiO(2--1). A horizontal arrow marks the expected velocity of
H$^{13}$CO$^+$(1--0) at each panel according to the measured centroid for the SiO(2--1) line (i.e., redshifted
$\sim$+321~km~s$^{-1}$ from the SiO line). The expected velocities for the four hyperfine lines of HCO are
marked by vertical arrows on the panel of GMC C. { Vertical $\pm1\sigma$ errorbars indicate the expected peak
  temperatures of the HCO hyperfine lines, fixing the HCO($J=3/2-1/2; F=2-1$) temperature from observations
  and assuming  optically thin emission for the four lines.
}
}
\label{figspe}
\end{figure*}

Fig.~\ref{figspe}
shows the SiO and H$^{13}$CO$^+$ spectra observed towards the positions of GMCs  A--to--E (the corresponding
gaussian fits are listed in Tab.~\ref{tabgau}).   As sketched in Fig.~\ref{figspe}, we report on the tentative
detection of two out of the four hyperfine line components of the $N_{K^-K^+}=1-0$ group of transitions of HCO
towards the position of GMC C. These correspond to the rest frequencies 86.671~GHz ($J=3/2-1/2, F=2-1$) and
86.777~GHz ($J=1/2-1/2, F=1-1$).  { The  HCO($F=2-1$)--to--H$^{13}$CO$^+$(1--0) intensity ratio inferred at
  GMC C is $\sim0.51\pm0.16$; comparable ratios ($\sim0.15-0.5$) were derived  by Garc\'{\i}a-Burillo et
  al.~(\cite{buri02}) for the $\sim$650~pc  nuclear disk of \object{M~82}, where the large HCO abundances
  ($X$(HCO)$\sim4\times10^{-10}$) indicate that   the whole inner disk can be viewed as a giant PDR.} The
tentative detection of  HCO towards C would be well accounted if UV fields are partly driving the chemistry of
the molecular clouds closest to the embedded star forming complex identified in the NIR by B\"oker et
al.~(\cite{boke97}) (see Garc\'{\i}a-Burillo et al.~\cite{buri02} for a discussion on the chemistry of the HCO
molecule).

\begin{table}[htp!]
\caption{Parameters of gaussian fits of the SiO(2--1), H$^{13}$CO$^+$(1--0) and HCO(1--0) lines of
Fig.~\ref{figspe}. Errors (in brackets) are 1$\sigma$ values.}
\label{tabgau}
\begin{tabular}{l@{\hspace{0.2cm}}l c@{$\;\;$}c r@{$\;\;$}c  c@{}c c@{}c}
\hline
\hline
\noalign{\smallskip}
&
\,\,{(1)}&
\multicolumn{2}{c}{(2)}&
\multicolumn{2}{c}{(3)}&
\multicolumn{2}{c}{(4)}&
\multicolumn{2}{c}{(5)}\\
{} & {Line}				&
\multicolumn{2}{c}{$T_\mathrm{peak}$}	&
\multicolumn{2}{c}{$I$}		      		&
\multicolumn{2}{c}{$ v_\mathrm{LSR}$}	 	&
\multicolumn{2}{c}{$ \Delta v_{1/2}$}    	 \\ 
\noalign{\smallskip}
		&			&
\multicolumn{2}{c}{(mK)}  		&			
\multicolumn{2}{c}{(K~km~s$^{-1}$)}	&
\multicolumn{2}{c}{(km~s$^{-1}$)} 	&
\multicolumn{2}{c}{(km~s$^{-1}$)}	 \\ 
\noalign{\smallskip}
\hline
            \noalign{\medskip}
                A      &
              SiO(2--1)& 27	& (6)	& 0.96	& (0.15)	& 18 	& (5)	& 34	& (6) \\
&$\rm H^{13}CO^+$(1--0)& 37	& (6)	& 1.37 	& (0.16) 	& 17    & (5) 	& 35 	& (5) \\
            \noalign{\smallskip}
\cline{2-10} 
  \noalign{\medskip}
                B      &
              SiO(2--1)& 22	& (5)	& 0.54 	& (0.10) 	& 17 	& (5)	& 23 	& (5) \\
&$\rm H^{13}CO^+$(1--0)& 43	& (5)	& 1.15 	& (0.12) 	& 17	& (5)	& 25	& (5) \\
            \noalign{\medskip}
\cline{2-10} 
  \noalign{\medskip}
                C      &
              SiO(2--1)& 48 	& (5)	& 2.26 	& (0.21)	& 43 	& (5)	& 45 	& (6) \\
&$\rm H^{13}CO^+$(1--0)& 42	& (5)	& 1.10 	& (0.14) 	& 42 	& (5)	& 25 	& (5) \\
&	    HCO(F=1--1)& 17	& (5)	& 0.58	& (0.16)	& 50	& (5)	& 32	& (9) \\
& 	    HCO(F=2--1)& 16	& (5)	& 0.56	& (0.16)	& 49	& (6)	& 32	& (9) \\
            \noalign{\medskip}
\cline{2-10} 
  \noalign{\medskip}
                D      &
              SiO(2--1)& 31	& (5)	& 1.61 	& (0.20) 	& 51	& (5)	& 48 	& (8) \\
&$\rm H^{13}CO^+$(1--0)& 20	& (5)	& 0.52 	& (0.12) 	& 52 	& (5)	& 24 	& (5) \\
            \noalign{\medskip}
\cline{2-10} 
  \noalign{\medskip}
                E      &
              SiO(2--1)& 42	& (6)	& 1.03 	& (0.11) 	& 13 	& (5) 	& 23 	& (5) \\
&$\rm H^{13}CO^+$(1--0)& 44	& (6)	& 1.05 	& (0.12) 	& 15 	& (5) 	& 22 	& (5) \\
            \noalign{\medskip}

\hline
\end{tabular}\\

\end{table}

 
\section{SiO fractional abundances} 
\label{secabu}

We have estimated the column densities { (${N}$)} of SiO and H$^{13}$CO$^+$ from the PdBI map of
\object{IC~342} using a Large Velocity Gradient (LVG) code. Our aim is to estimate the abundance of SiO
relative to H$_2$ (X(SiO)) inferred here from the SiO--to--H$^{13}$CO$^+$ column density ratio ($R_N$).  
  Values of   $N$(SiO) and $N$(H$^{13}$CO$^+$)  are derived from measured integrated intensities,   
 assuming a plausible range of physical conditions for the gas. Given that the emission of both species is
 optically thin, $R_N$ is proportional to the  SiO/H$^{13}$CO$^+$ intensity ratio ($R_I$). Since SiO and
 H$^{13}$CO$^+$ have similar dipole moments and the observed transitions have comparable upper state energies,
 it is reasonable to assume the same physical conditions for both species.  We have run five LVG models
 covering a range of gas densities ($n$(H$_2$)) from $10^4$~cm$^{-3}$ to 10$^6$~cm$^{-3}$. The explored
 interval encompasses the total range of molecular gas densities determined in the GMCs of \object{IC~342}
 from multitransition studies of CO and HCN  (e.g., see Schulz et al.~\cite{schu01}). As gas kinetic
 temperature, we adopt a value $T_\mathrm{K}=50$~K (Downes et al.~\cite{down92}).  The value of $T_\mathrm{K}$
 is not critical in the estimate of $R_N$ within the range of $n$(H$_2$) explored in these calculations: the
 inferred column density ratios are similar within the temperature interval  20~K$\leq T_\mathrm{K}\leq$80~K
 and $R_N$ increases, at most, by 40$\%$ if $T_\mathrm{K}$ is lowered to 10~K. 

\begin{figure}[thp!]
\includegraphics[angle=-90,width=\hsize]{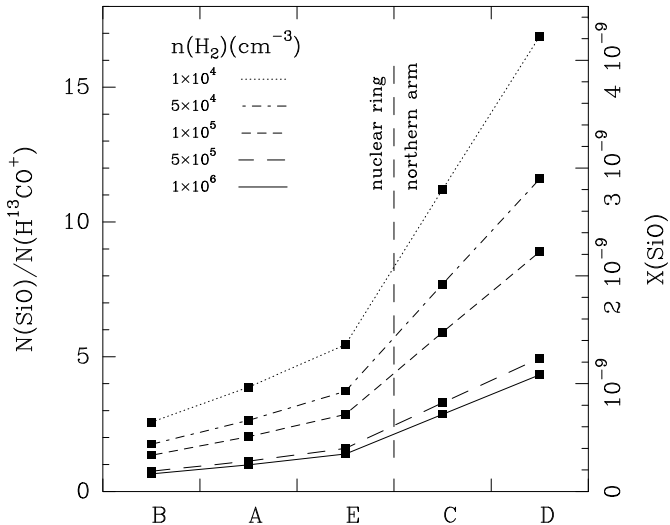}
\caption{
SiO-to-H$^{13}$CO$^{+}$ column density ratios and SiO abundances in GMCs A--to--E derived from a LVG
calculation, assuming $T_\mathrm{K}$=50~K and five different values of $n$(H$_2$) from $10^4$ to
$10^6$~cm$^{-3}$. GMCs are ordered along the X-axis by increasing X(SiO).
}   
\label{figabu} 
\end{figure}

We show in Fig.~\ref{figabu} the values estimated for $R_N$ towards GMCs A--to--E for $T_\mathrm{K}$=50~K
particularized for the different $n$(H$_2$) values. On average, $R_N$ is seen to increase by a factor of
$\sim4$ when $n$(H$_2$) is lowered from 10$^6$~cm$^{-3}$ to 10$^4$~cm$^{-3}$. 
The fractional abundance of SiO is inferred from $R_N$ assuming a {\it standard} abundance for H$^{13}$CO$^+$.
The assumption of a standard value for X(H$^{13}$CO$^+$) is supported by observations of molecular clouds in our Galaxy.
Contrary to SiO, for which measured abundances are seen to differ by several orders of magnitude between
quiescent clouds and shocked regions, H$^{13}$CO$^+$ shows a fairly stable abundance in a large variety of
physical and chemical environments (see discussion in Garc\'{\i}a-Burillo et al.~\cite{buri00}).  
Here we adopt X(H$^{13}$CO$^+$)=2.5$\times$10$^{-10}$; this corresponds to a typical abundance of the 
main isotope X(H$^{12}$CO$^+$)=$10^{-8}$ and to an isotopic ratio [$^{12}$C]/[$^{13}$C]$\simeq$40 (Henkel 
et al.~\cite{henk98}). 

The estimated abundance of SiO is $>2\times10^{-10}$ for all GMCs, i.e., at least two orders of
magnitude larger than the typical SiO abundances of Galactic quiescent clouds (Mart\'{\i}n-Pintado, Bachiller
\& Fuente~\cite{mart92}). This lower limit on the beam-averaged value of X(SiO) measured here on GMC-like
scales  { (${\sim80}$~pc)} indicates that shock chemistry is at work in the inner { ${r\sim320}$~pc} disk of
\object{IC~342}. Most remarkably, we see an overall N-S gradient in the value estimated for X(SiO). On
average, the abundance of SiO is nearly one order of magnitude larger in the northern spiral arm (e.g.,
$\sim$1--4$\times$10$^{-9}$ at GMC D) than in the nuclear ring  (e.g., $\sim$2--7$\times$10$^{-10}$ at GMC B)
within the explored range of densities. Schulz et al. (\cite{schu01}) have estimated the average densities of
GMCs A--to--E based on a multitransition  study of HCN and their isotopes. The results of this study indicate
that average densities are a factor 2 larger in the nuclear ring GMCs compared to that measured in the
northern spiral arm. This would imply that the N-S gradient on the value of X(SiO) may be even larger than
estimated above. Moreover, the overall mass budget of  \object{IC~342} is seen to be heavily
weighted by molecular gas with typical densities $\sim10^3-10^4$~cm$^{-3}$  (Downes et al.~\cite{down92};
Israel \& Baas~\cite{isra03}). Should this diffuse gas partly contribute  to the emission of both species this
would imply that the SiO abundances may have to be  boosted for all the  GMCs (see Fig.~\ref{figabu})
compared to the values reported above, thus reinforcing the case for shock chemistry.    

The beam-averaged values of X(SiO) derived above are necessarily lower limits to the real abundance attained
by SiO in the fraction of dense molecular gas which is being processed by shocks
in \object{IC~342} (hereafter X(SiO)$|_{shock}$). If we denote by $f_{shock}$ the {\it a priori} unknown
fraction of shocked dense gas, X(SiO) can 
be formally factorized as:   

\begin{equation} 
X\mathrm{(SiO)}=f_{shock}\times X\mathrm{(SiO)|_{shock}} 
\label{equsio}
\end{equation}

The notable N--S gradient found in the derived value of X(SiO) between the nuclear ring and the northern spiral arm
could be accounted by two extreme scenarios: first, an order of magnitude change in $f_{shock}$ between the two 
regions, or alternatively, a similar change in X(SiO)$|_{shock}$. While the first scenario would imply that
the typical column densities of shocked gas in the northern spiral arm are $\sim6-7$ times larger than in the
nuclear ring, the second scenario would call for a notably different shock velocity regime ($v_{shock}$) in
the two regions. In this case $v_{shock}$ should be significantly larger in the northern spiral arm. As is
discussed in Sect.~\ref{secwav}, finding the right scenario between the two proposed above is key to shed
light on the nature of the driving mechanism of shocks in \object{IC~342}. 
In Sect.~\ref{seceff} we compare the emission of SiO in \object{IC~342} with that of
CH$_3$OH, another molecular shock tracer. In particular, we use the SiO--to--CH$_3$OH ratio derived for
\object{IC~342} to explore the origin of shocks in this galaxy. 


\section{Molecular shock chemistry in \object{IC~342}}
\label{seceff}

\subsection{Tracers of shocks in molecular gas}
\label{subche}

The significant enhancement of SiO in gas phase is considered to be an indication that shock
chemistry is at work in molecular gas (Mart\'in-Pintado et al.~\cite{mart92}).  The injection of Si-bearing
material from dust grains into the gas phase, either through sputtering or  grain--grain collisions, can
explain the measured abundances of this molecule in shocked regions (Schilke et al. \cite{schi97}; Caselli et
al. \cite{case97}).  Shocks are often invoked to account for the large abundances of other molecular species
measured in bipolar outflows. This is  the case of CH$_3$OH (Bachiller et al. \cite{bach95}). Although
qualitatively similar, shocks characterized by different velocity regimes are expected to process to a
different extent dust grains in molecular gas. {\it  Fast} shocks ($v_{shock}>$15--20~km~s$^{-1}$) can destroy
the grain cores, liberating refractory elements to the gas phase (Schilke et al. \cite{schi97}; Caselli et
al. \cite{case97}). In contrast, while {\it slow} shocks ($v_{shock}<$10--15~km~s$^{-1}$) are not able to
destroy the grain cores, they can heavily process the icy grain mantles. The assumed different location of
Si-bearing material (cores) and solid-phase CH$_3$OH (mantles) in dust grains makes of SiO and  CH$_3$OH good
tracers of {\it fast} and {\it slow} shocks, respectively. Furthermore, for velocities above
$\sim$10--15~km~s$^{-1}$ shocks could destroy the molecules in gas-phase of volatile species such as CH$_3$OH
(Garay et al.  \cite{gara00}; J{\o}rgensen et al.~\cite{jorg04}). The dissociation of SiO by shocks would
require velocities $\gtrsim$50-60~km~s$^{-1}$ (i.e. J shocks), however.

If we consider both processes, i.e., the injection of grain material and the disruption of molecules in gas phase, we can 
conclude that an increase in the typical velocity regime of shocks ($v_{shock}$) will certainly favour an 
enhancement of the abundance of SiO in the shocked gas (X(SiO)$|_{shock}$). This will be at the expense of
increasing the SiO--to--CH$_3$OH abundance ratio in molecular gas. Therefore, a variation in the
SiO--to--CH$_3$OH intensity ratio can be taken as a evidence for a change in $v_{shock}$. A quantitative
comparison of the SiO and CH$_3$OH maps of \object{IC~342} could thus help to 
discern if the typical shock velocity regime changes across the galaxy disk.

\subsection{Tracers of shocks in \object{IC~342}: SiO and CH$_3$OH}
\label{submet} 

\begin{figure}
\includegraphics[angle=0,width=\hsize]{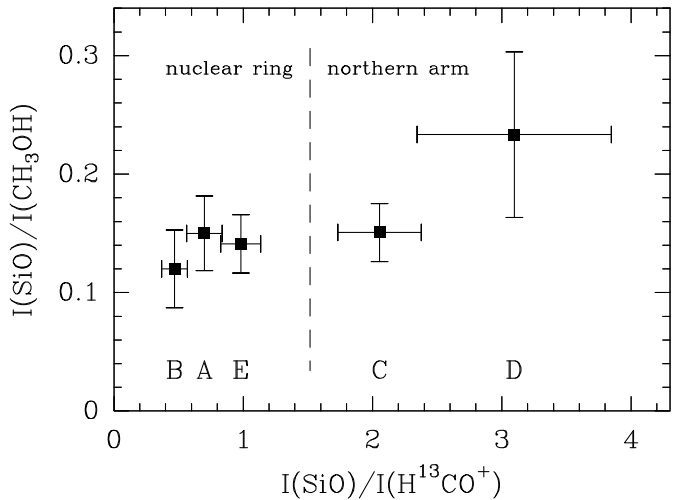}
\caption{
We represent the SiO--to--CH$_3$OH integrated intensity ratio as a function of the SiO--to--H$^{13}$CO$^{+}$
ratio (R$_I$) in GMCs A--to--E. CH$_3$OH data have been taken from Meier \& Turner (\cite{meie05}).
}  
\label{figpar}
\end{figure}

Meier \& Turner (\cite{meie05}) used the OVRO interferometer to map the emission of the
2$_\mathrm{k}$--1$_\mathrm{k}$ line of CH$_3$OH in \object{IC~342} with a resolution similar to that of the
SiO map. This spatial resolution is comparable to the typical GMC-scales { (${\sim80}$~pc)}. We have compared
the emission of SiO(2--1)  to that of CH$_3$OH(2$_\mathrm{k}$--1$_\mathrm{k}$) in order to derive the
SiO--to--CH$_3$OH intensity ratio in GMCs A--to--E. Fig.~\ref{figpar} represents these ratios as a function of
$R_I$ for the nuclear ring and northern spiral arm GMCs. As is shown in Fig.~\ref{figpar}, the reported nearly
order of magnitude change in $R_I$ is not corresponded with a similar change in the I(SiO)/I(CH$_3$OH) ratio
between the northern spiral arm and the nuclear ring. The  I(SiO)/I(CH$_3$OH) ratio is fairly constant and close to
$\sim$0.14\footnote{
Assuming the CH$_3$OH column densities given by Meier et al.~(\cite{meie05})
these intensity ratios translate into SiO--CH$_3$OH column density ratios of $\sim0.01-0.04$ for
$n\gtrsim5\times10^4$~cm$^{-3}$ ($\sim3$ times larger if $n\sim10^4$~cm$^{-3}$).
} 
 for all GMCs, though we find tentative evidence for a larger value in GMC D ($\sim$0.24$\pm$0.07).  Taken
 together these results indicate that, at first order, the N-S gradient measured in $R_I$ between the
 northern spiral arm and the nuclear ring can be mostly attributed to a variation of $f_{shock}$.

\begin{figure}
\includegraphics[angle=0,width=\hsize]{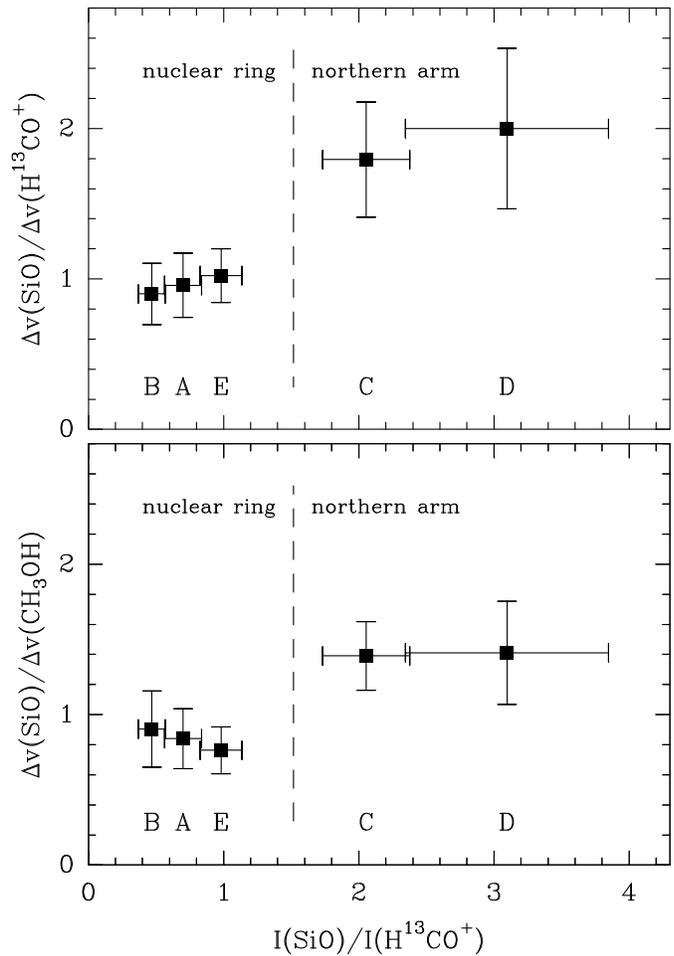}
\caption{
SiO--to--H$^{13}$CO$^+$ and SiO--to--CH$_3$OH velocity width ratios measured for GMCs A--to--E. These
  ratios are represented as a function of the SiO--to--H$^{13}$CO$^+$ integrated intensity ratio, $R_I$.
} 
\label{figdvr}
\end{figure}

The line ratios above are derived from velocity-integrated intensities and are beam-averaged on scales which
are typical of GMC-like units at the distance of \object{IC~342}.  However, the analysis of the line profiles
of SiO, CH$_3$OH and H$^{13}$CO$^+$ can provide information on the shock velocity regime on scales smaller
than the beam. Fig.~\ref{figdvr} displays the  SiO--to--H$^{13}$CO$^+$ and SiO--to--CH$_3$OH velocity--width
ratios derived for GMCs A--to--E. These ratios show a different behaviour in the nuclear ring and in the
northern spiral arm. In the ring, linewidths and velocity centroids for SiO, H$^{13}$CO$^+$ and CH$_3$OH are
virtually identical. In contrast, SiO lines are a factor of $\sim$2 wider than that of H$^{13}$CO$^+$ in the
northern spiral arm, as reported in Sect.~\ref{subkin}. CH$_3$OH lines  represent a case intermediate between
these two extremes: SiO lines are a factor of $\sim$1.4 wider than that of CH$_3$OH in the spiral arm. The
differences between SiO, CH$_3$OH and H$^{13}$CO$^+$ are evident in the  linewidths, but velocity centroids
are the same within the errors. { As discussed in Sect.~\ref{secwav}, this suggests that the
  \emph{apparent turbulence} of shocked molecular gas is enhanced compared to the more quiescent gas.}

While it seems that the enhancement of $f_{shock}$ in the northern spiral arm explains the bulk of the reported
increase of X(SiO) in this region, the observed differences in the line profiles of SiO and CH$_3$OH  
suggest that a fraction of the shocked gas in the arm presents a higher X(SiO)$|_{shock}$. This implies 
that $v_{shock}$ would be a factor of $\sim$2 larger in the spiral arm region.


\section{SiO emission in Galaxies}
\label{secori}

There is ample observational evidence that SiO thermal emission can be locally enhanced in star forming
molecular clouds of our Galaxy (Mart\'{\i}n-Pintado et  al. \cite{mart92}; Bachiller et
al. \cite{bach01}). More recently, SiO has been revealed to be a tracer of shock chemistry also in galaxy
nuclei, including our own Galaxy (Sage \& Ziurys~\cite{sage95}; Mart\'{\i}n-Pintado et al.~\cite{mart97};
Garc\'{\i}a-Burillo et al.~\cite{buri00, buri01b}; Usero et al.~\cite{user04}).  We revise below the various
driving mechanisms which have been proposed to explain the onset of shock chemistry  in the Galaxy and in
galaxies in general, and study their applicability to the nucleus of \object{IC~342}.

\subsection{SiO emission in the disk of our Galaxy}
\label{subgdi}

  SiO thermal emission is observed towards star-forming clouds in the disk of our Galaxy. In particular, the strongest 
SiO emitters are the bipolar outflows located around protostellar objects (Mart\'{\i}n-Pintado et  al. \cite{mart92};
Bachiller et al. \cite{bach01}). In the first stages of the star formation process, bipolar outflows 
interact with the ambient molecular gas, inducing molecular shocks which are able to increase the abundances of some 
molecular species (like SiO and CH$_3$OH) by several orders of magnitude with respect to quiescent gas 
(Bachiller et al. \cite{bach95,bach97}).

The abundance of SiO in \object{IC~342} reaches the largest value 
in the northern spiral arm (Sect.~\ref{secabu}), where the evidence of ongoing star formation is
scarce (Sect.~\ref{subcon}). However, it could be argued 
that SiO emission  in \object{IC~342} is probing the deeply embedded phase of a young star formation
episode. The episode would be spread on scales of $\sim$a few 100~pc and would not visible in H$\alpha$ or
thermal radio-continuum. Although with these tight constraints this explanation is unlikely, we can discard it
on more quantitative grounds comparing the SiO emission in \object{IC~342} with a subset of bipolar outflows
of the Galaxy (Tab.~\ref{tabyso}). For the reasons explained below, we have purposely selected outflows  for
which there are maps of both SiO(2--1) and  CH$_3$OH(2$_\mathrm{k}$--1$_\mathrm{k}$) available in the
literature. Furthermore,  since SiO luminosities are larger for more massive and younger objects, the outflows
have been chosen to cover a wide range in mass and age of the protostars. It is assumed  that the largest
SiO--to--CH$_3$OH luminosity ratios correspond to the less evolved objects (Bergin et al.\cite{berg98}).

\begin{table*}
\caption{
SiO(2--1) and CH$_3$OH(2$_\mathrm{k}$--1$_\mathrm{k}$) luminosities of a sample of bipolar outflows. {\sl
  Col.~2}: mass of the class 0 object. {\sl Col.~3}: luminosity of the class 0
object. {\sl Col.~4}: SiO luminosity integrated within the SiO emitting region of the outflow. {\sl
  Col.~5}: SiO--to--CH$_3$OH luminosity ratio. {\sl Col.~6}: surface density of outflows
required to obtain the mean $I$(SiO) in the northern spiral arm+nuclear ring of \object{IC~342}. {\sl Col.~7}:
References: \textit{a}: Bachiller et al. (\cite{bach01}); \textit{b}: Gueth et al. (\cite{guet97});
\textit{c}: Fuente et al. (\cite{fuen05b}); \textit{d}: Codella \& Bachiller et al. (\cite{code99}).  
}  
\label{tabyso} 
\begin{tabular}{llrc ccl}
\hline
\hline
\noalign{\smallskip}
\phantom{So}{(1)}&
\multicolumn{1}{c}{(2)}&
\multicolumn{1}{c}{(3)}&
\multicolumn{1}{c}{(4)}&
\multicolumn{1}{c}{(5)}&
\multicolumn{1}{c}{(6)}&
\multicolumn{1}{c}{(7)}\\
{Source} &\multicolumn{1}{c}{M$_{\star}$} &\multicolumn{1}{c}{L$_\star$} & L$_\mathrm{SiO}$ &
$\mathrm{L_{SiO}/L_{CH_3OH}}$ &\multicolumn{1}{c}{$N_\mathrm{outf}$}&Ref.\\    
\noalign{\smallskip}
	 &(M$_{\sun}$) 	& (L$_{\sun}$)	& (K~km~s$^{-1}$~pc$^2$) & &\multicolumn{1}{c}{(outflows~pc$^{-2}$)}\\ 
\noalign{\smallskip}
\hline
\noalign{\medskip}
\object{L~1157}(blue lobe)				
                           &0.2	& 11	& 0.19	& 0.38	&\phantom{1}4.9 	
& \textit{a,b}\\  

\object{NGC~7129-FIRS2}                                 				
                          & 5& 500                   & 	& 	&    	&  \textit{c}\\
\multicolumn{1}{r}{outflow 1}&	&                         & 0.21	& 0.14 	&\phantom{1}4.4 \\                        
\multicolumn{1}{r}{outflow 2}&       	&	& 0.08	& 0.10                  & 11.6\\
\object{CB3}								
                           & 4	& 930	& 2.37	& 0.45	&\phantom{1}0.4 
& \textit{d} \\
\noalign{\medskip}
	    
\hline
\end{tabular}

\end{table*}

The mean SiO intensity of the northern spiral arm and the nuclear ring of \object{IC~342}, derived inside the region defined
by the lowest contour of Fig.~\ref{figint}, is $\sim$0.93~K~km~s$^{-1}$. We have then inferred the surface
density of  outflows which would be required to reproduce the observed SiO intensity in \object{IC~342}.  As
is shown in Tab.~\ref{tabyso}, this number density of outflows range from $\sim$0.4 to
$\sim$12~outflows~pc$^{-2}$.  Though in star-forming regions of our Galaxy like OMC2/3 or \object{NGC~2068},
surface densities of a few outflows~pc$^{-2}$ have been reported on scales of 1--2~pc$^2$ (Reipurth et
al.~\cite{reip99}; Mitchell et  al.~\cite{mitc01}), it  is very unlikely that these average surface densities
can be attained in the SiO disk of \object{IC~342}, which is several hundred pc$^2$ in extent.  

Furthermore the star forming  rate (SFR) of \object{IC~342} poses tight constraints on the upper limit to the
expected density of outflows. The SFR in a galaxy can be easily estimated from the FIR luminosity (Kennicutt
\cite{kenn98}). In the case of \object{IC~342},  { L${_\mathrm{FIR}=1.5\times10^9}$~L$_{\sun}$ in the central
$r\sim15\arcsec$  (Becklin et  al. \cite{beck80}; scaled to D=3.3~Mpc)}. Assuming that the northern spiral arm and
nuclear ring are the main contributors to the SFR of \object{IC~342}, we estimate for this region a SFR
density of $3\times10^{-6}$~M$_{\sun}$~yr$^{-1}$~pc$^{-2}$. From the SFR density above, we can then derive an
upper limit to the surface density of outflows, assuming a Salpeter law for the IMF
($\mathrm{d}\log(N)/\mathrm{d}\log(m)=-2.35$ over $m=0.1-100$~M$_{\sun}$) and a timescale for the pre-stellar 
phase of $\sim10^4$~yr. The derived upper limit to the total density of outflows provided by the SFR in
\object{IC~342} is $\sim9\times10^{-2}$~outflows~pc$^{-2}$. Even for outflows like \object{CB3} (or any other SiO
luminous YSO), the required density would be $\sim$5 times larger than that provided by the SFR of
\object{IC~342}.  Moreover,  massive young bipolar outflows like \object{CB3} should be the minority among
YSOs: for a standard Salpeter IMF, only 4\% of the objects would have masses above 1~M$_ {\sun}$.  In
addition, the observed SiO--to-CH$_3$OH average ratio of \object{IC~342} ($\sim0.14$; see Fig.~\ref{figpar})
is similar to YSOs which are much more evolved than \object{CB3}.

In summary, we can discard the interpretation of the  large-scale shocks in \object{IC~342} in terms of a
collection of YSOs associated with an embedded star formation episode in the inner { ${r\sim320}$~pc} disk of the
galaxy.

\subsection{SiO emission in galactic nuclei}

\subsubsection{The center of our Galaxy}
\label{subgce}

The nucleus of our Galaxy shows widespread SiO emission that, in contrast to that observed in the Galactic 
Disk, is not related to recent star formation. The first large-scale SiO observations of the Galactic Center (GC) of
Mart\'{\i}n-Pintado et al.~(\cite{mart97}) detected the emission of the SiO(1--0) line in a 
$\sim150$~pc--diameter circumnuclear disk. In a later paper, H\"uttemeister et al. (\cite{huet98}) detected the  SiO(2--1) emission
in 32 GC molecular clouds located inside a $\sim$300~pc--diameter disk which extends from \object{Sgr~C} to a position at slightly 
higher positive longitudes than \object{Sgr~B2}. High SiO abundances are derived for these clouds ($\sim$
several $10^{-10}$ - several $10^{-9}$), indicative of shock chemistry. This scenario has received further support from recent
observations revealing the complex alcohol chemistry of GC clouds (Mart\'{\i}n-Pintado et al. \cite{mart01};
Requena-Torres et al.~2005, in  prep.). In these clouds the large abundances of ethanol (C$_2$H$_5$OH) and CH$_3$OH evidence
the erosion of dust grain mantles on large scales.

Besides the long reported similarities between our Galaxy and \object{IC~342} (e.g., Downes et al. \cite{down92}),
the inner { few hundred~pc} in the two galaxies seem to be the scenario of large-scale molecular shocks
leading to dust grain processing. Furthermore, the efficiency of shocks appear to be comparable in both
galaxies. Despite the different spatial resolution of SiO observations in \object{IC~342} { ($\sim$80~pc)} and in
our Galaxy ($\sim$2~pc), the derived SiO abundances are similar in the two objects. Moreover, the
SiO-to-CH$_3$OH abundance ratios are also similar in the GC clouds ($\sim0.01-0.03$)\footnote{Calculated from
  SiO and C$_2$H$_5$OH column densities measured in GC clouds (Mart\'{\i}n-Pintado et al. \cite{mart97,mart01}),
  assuming a constant C$_2$H$_5$OH--to--CH$_3$OH abundance ratio of $\sim$0.05  (Requena-Torres et al.~2005, in
    prep.).} and in \object{IC~342} ($\sim0.01-0.04$; see Sect.~\ref{submet}).

These similarities taken together, we can hypothesize that the mechanism explaining the onset of
large-scale molecular shocks in the GC and in \object{IC~342} is likely to be the same.
Shocks identified in GC clouds, also unrelated to ongoing star formation, have been attributed different causes, however: 
the interaction with GC non-thermal filaments, with supernovae remnants or with the expansion 
bubbles created by Wolf-Rayet stars have been discussed by Mart\'{\i}n-Pintado et al.~(\cite{mart97}). 
H\"uttemeister et al. (\cite{huet98}), using observations that extended over a wider region in the GC, found
the largest SiO abundances  where the likelihood of cloud-cloud collisions, induced by the Galactic bar
potential, is the highest.  
As discussed in Sects.~\ref{subexg} and \ref{secwav}, the latter scenario is the preferred one in \object{IC~342}.

\subsubsection{The nuclei of external galaxies} 
\label{subexg}

\begin{table*}
\caption{
SiO emission in external galaxies. {\sl Col.~2}: SiO abundance averaged within $28\arcsec$ (from
observations with the IRAM 30~m telescope); { the averaging scale is $\sim0.5$~kpc, for  \object{NGC~253},
  \object{IC~342} and \object{M~82}, and $\sim2$~kpc for \object{NGC~1068}.} {\sl
  Col.~3:} components resolved with the PdBI contributing to the global SiO emission; {\sl Col.~4}: SiO
abundance in the resolved components. {\sl Col.~5}: mechanism proposed to be  enhancing the SiO abundance;
underlying physical process, in brackets (DW stands for `density waves'). 
{Col.~6}:  References: {\it a}: Garc\'{\i}a-Burillo et al. (\cite{buri00}); {\it b}: this
  work;   {\it c}: Garc\'{\i}a-Burillo et al. (\cite{buri01b});  {\it d}: Usero et al. (\cite{user04}); {\it} e:
  Mart\'{\i}n-Pintado et al.~(2005, in prep.).  
}
\label{tabgal}
\begin{tabular}{lclcll}
\hline
\hline
\noalign{\smallskip}
\multicolumn{1}{c}{(1)}&\multicolumn{1}{c}{(2)}&\multicolumn{1}{c}{(3)}&\multicolumn{1}{c}{(4)}&\phantom{SiO
  ch}(5)&\multicolumn{1}{c}{(6)}\\    
Galaxy &\multicolumn{1}{c}{$\langle X\mathrm{(SiO)}\rangle_\mathrm{30m}$}& Resolved components &  
\multicolumn{1}{c}{X(SiO)}&\multicolumn{1}{l}{SiO Chemistry}&Ref.\\     
 & \multicolumn{1}{c}{$(10^{-10})$} & &\multicolumn{1}{c}{$(10^{-10})$}\\ 
\noalign{\smallskip}
\hline
\noalign{\medskip}
\object{NGC~253}&$1-3$ & & & & \it{a,e}\\
& &$r\sim60$~pc inner ring &$1-2$ & shocks[YSOs / DW]\\ 
& &$r\sim300$~pc outer ring &$5-15$& shocks[DW]\\  
\noalign{\smallskip}
\cline{3-6}
\noalign{\smallskip}
\object{IC~342}&$3-7$& & & & \it{b,e}\\
& & $r\sim50$~pc ring & $2-7$& shocks[DW]\\
& & 150~pc length arm & $10-40$& shocks[DW]\\
\noalign{\smallskip}
\cline{3-6}
\noalign{\smallskip}
\object{M~82}&0.5 & & & & \it{c,e}\\
& &$r\sim75$~pc shell & $0.4-1$   &shocks[galactic outflow]\\         
& &500~pc length chimney  &$2-3.5$ &shocks[galactic outflow]\\    
\noalign{\smallskip}
\cline{3-6}
\noalign{\smallskip}
\object{NGC~1068}&7 & & & & \it{d,e}\\
& &$r\sim1-1.5$~kpc ring &$\sim$a~few & shocks[YSOs / DW]\\  
& &$r\sim200$~pc disk &$\sim50$ & XDR[AGN] / shocks[DW]\\
\noalign{\medskip}
	    
\hline
\end{tabular}
\end{table*}

Mauersberger \& Henkel (\cite{maue91}) detected the emission of SiO(2--1) in the starburst galaxy
\object{NGC~253}, using the 30~m IRAM telescope  with a resolution of $\sim$28$\arcsec$.  This detection, the
first reported for SiO outside the Milky Way, was followed by a 9-galaxy survey made with the NRAO 12~m
antenna by Sage \& Ziurys (\cite{sage95}) with $\sim$67$\arcsec$ spatial resolution (equivalent to
$\sim$0.5--2~kpc). In this survey, where 5 galaxies were detected in SiO,  Sage \& Ziurys (\cite{sage95})
found no correlation between the abundance of SiO, characterized in their work by the SiO/N$_2$H$^+$ ratio,
and the efficiency of star formation (given by the SFR per unit dynamical mass, i.e., SFR/M$_{dyn}$). This
surprising result was at odds with the classical framework where shock chemistry is driven by YSOs in star
forming regions of the Galactic disk. The global SiO abundance measured on 0.5--1~kpc--scales is
seen to vary significantly among starburst galaxies: it can reach $\sim$10$^{-9}$ in \object{IC~342}, whereas
it is 1/20 of this value in \object{M~82} (see Table~\ref{tabgal} and Mart\'{\i}n-Pintado et al.~2005, in
prep.).

The occurrence of large-scale molecular shocks may arise at different stages during the typical lifetime of a
starburst episode (Rieke et al.~\cite{riek88}; Garc\'{\i}a-Burillo \& Mart\'{\i}n-Pintado~\cite{buri01a}).  
In the pre-starburst phase ({\bf I}), density wave instabilities induce
gravitational torques and drive the infall of large amounts of gas towards the nucleus. {\it Large-scale}
shocks may be at work related to an enhanced compression of gas and an increased rate of cloud-cloud
collisions in the potential wells of spiral arms and/or bars. Once the first massive stars are formed in a
second phase ({\bf II}), bipolar outflows can produce {\it locally} molecular shocks in YSOs. In a later stage,
corresponding to an evolved starburst, the elevated rate of SN explosions may lead to the disruption of the
disk during the expansion of the so-called {\it hot bubble}. Episodes of mass injection from the disk into the
halo could be accompanied by molecular shocks.  

The advent of high-resolution SiO images has been key to help discern the different sources of shock chemistry
in external galaxies and  thus identify an evolutionary path along the starburst sequence depicted above (see
Table~\ref{tabgal} and references therein). In particular, the puzzling result issued from first single-dish
SiO surveys starts to be understood when observations allow us to zoom in on molecular galaxy disks on scales
$\lesssim$100~pc:   

\begin{itemize}

\item 
Large-scale shocks near the outer Inner Lindbland Resonance (oILR) of the \object{NGC~253} stellar bar can account
for the outer SiO disk of this galaxy, which extends well beyond the nuclear starburst. The outer disk SiO
emission gives away the {\it pre-starburst phase} ({\bf I}) in  \object{NGC~253} (Garc\'{\i}a-Burillo et
al.~\cite{buri00}). More recently, the detection of SO$_2$, NS and NO emission and the analysis of the sulfur
chemistry in \object{NGC~253} have confirmed that shocks are at work in the nucleus of this galaxy 
(Mart\'{\i}n et al.~\cite{mart03, mart05}). 

 \item
Outflows driven by YSOs during phase {\bf II} can partly explain the measured average abundances of SiO in the
inner disk of \object{NGC~253} (Garc\'{\i}a-Burillo et al.~\cite{buri00}). However, the bulk of the SiO
emission likely stems from molecular cloud shocks in the inner Inner Lindbland Resonance (iILR) of the
\object{NGC~253} stellar bar.  

\item 

The emission of SiO extends noticeably out of the galaxy plane in \object{M~82}, tracing the disk-halo
interface where episodes of mass injection from the disk are building up the gaseous halo (phase {\bf III})
(Garc\'{\i}a-Burillo et al.~\cite{buri01b}).  The PdBI maps of \object{M~82} made in SiO and HCO illustrate how two
different gas chemistry scenarios can be simultaneously at play in the same galaxy though at different
locations: shocks in the disk-halo interface and PDR chemistry in the galaxy disk which hosts an evolved
starburst (Garc\'{\i}a-Burillo et al.~\cite{buri02}; Fuente et al.~\cite{fuen05a}). 
	
\end{itemize}

The case of the Seyfert 2 galaxy \object{NGC~1068} has been studied by Usero et al.~(\cite{user04}) who
estimate a large abundance of SiO in the  $r\sim$~200~pc CND of the galaxy (X(SiO)$\sim$a few~10$^{-9}$). The
enhancement of SiO is attributed by Usero et al.~(\cite{user04}) to the evaporation of very small  ($10~\AA$)
silicate grains (VSG) by X-rays. Alternatively, silicon chemistry could be also driven by pre-starburst shocks
related with the density wave-resonances of the CND (Garc\'{\i}a-Burillo et al.~2005, in prep.) 

As extensively argued in Sect.~\ref{subgce}, the SiO emission in \object{IC~342} cannot be explained by
ongoing star forming activity (phase {\bf II}). The \object{M~82} scenario (phase {\bf III}) can be also ruled out in
\object{IC~342}. SiO emission in the disk of \object{IC~342} extends well beyond the distribution of
supernovae remnants (SNRs) (Condon et al. \cite{cond82}; Bregman et al. \cite{breg93}). Furthermore, the low
supernovae rate of \object{IC~342} ({ ${<0.04}$~yr$^{-1}$ for D=3.3~Mpc}; Condon et al. \cite{cond82}) yields a
{ ${\sim 70}$ times} smaller energy deposition by the SN of \object{IC~342} compared to \object{M~82}.  
Finally, X-rays are not expected to be a dominating agent in the chemistry of molecular gas in \object{IC~342}
either. Nearly $\sim$35\% of the observed  hard X-ray emission in \object{IC~342} (2~keV$\leq$~E~$\leq$10~keV)
comes from a circumnuclear disk of $r\sim8\arcsec$ (Bauer, Brandt \& Lehmer~\cite{baue03}). The hard X-ray
luminosity of this source is three orders of magnitude lower than that of \object{NGC~1068} (Ogle et
al. \cite{ogle03}), however. Furthermore, the 6.4~keV Fe K$\alpha$ line, which probes the processing of
neutral gas by X-rays, is absent from the spectrum of \object{IC~342}.

The exclusion of all the alternative explanations leads us to conclude that the large-scale shocks identified
in the inner { ${r\sim320}$~pc} of \object{IC~342} arise in the
  pre-starburst phase ({\bf I}) { (see Sect.~\ref{seccon})}. The following Section 
discusses the efficiency of density--waves at producing shocks in molecular gas in \object{IC~342}.

\section{Density waves and shocks in \object{IC~342}}
\label{secwav} 

The \object{IC~342} bar shapes the distribution and kinematics of molecular gas in the
central { ${r\sim320}$~pc}  of the galaxy (Turner \& Hurt \cite{turn92}; Schinnerer et al.~\cite{schi03}; Meier \&
Turner~\cite{meie05}; this work).  The  spiral-like morphology of the molecular disk and the
detection of non-circular motions ($\gtrsim 50-60$~km~s$^{-1}$ in the northern spiral arm, deprojected onto
the galaxy plane) are reminiscent of the typical bar-driven dynamics. The SiO abundances measured in the inner
molecular disk of  \object{IC~342} proves unambiguously that the bar is producing large-scale molecular
shocks. However, the detection of SiO  emission constrains the velocity regime of shocks to lie between
$\gtrsim15-20$~km~s$^{-1}$ (for grain cores  to be significantly disrupted) and $\lesssim50-60$~km~s$^{-1}$
(to prevent dissociation of SiO molecules), i.e., a velocity range which lies
significantly below the lower limit set to the non-circular motions measured across the \object{IC~342}
bar. Furthermore, the detection of CH$_3$OH emission across the bar suggests that the shocked molecular gas
emitting in CH$_3$OH cannot be characterized by $v_{shock}\gtrsim 50-60$~km~s$^{-1}$. These observational
constraints imply that the input kinetic energy provided by streaming motions must  be first dissipated making
room for a lower velocity regime that corresponds to the emission of the molecular shock tracers  observed
across the \object{IC~342} bar.

Large-scale shocks driven by density waves have been long predicted by numerical simulations of spiral/barred
galaxies, made following either hydrodynamical schemes (Roberts \cite{robe69}; Athanassoula \cite{atha92}) or
ballistic ones (Casoli \& Combes \cite{caso82}; Combes \& Gerin \cite{comb85}). In ballistic models, which
likely provide a more realistic representation of the clumpy dense ISM in galaxies, molecular shocks should
arise subsequently after cloud-cloud collisions. The number of collision events is enhanced by an increase of
orbit crowding along the potential well of the bar. As noted in Sects.~\ref{subkin} and \ref{submet}, SiO, 
H$^{13}$CO$^+$ and CH$_3$OH lines have all similar velocity centroids, but different linewidths over the spiral 
arm region. The largest widths correspond to SiO (whose lines are a factor of 2 larger than that of 
H$^{13}$CO$^+$), with CH$_3$OH representing an intermediate case. This suggests that the {\sl apparent
  turbulence} of shocked molecular gas, traced by SiO and CH$_3$OH, is enhanced compared to the state of the
more quiescent dense gas medium (traced by H$^{13}$CO$^+$).  
This effect, especially relevant in the northern spiral arm region (see Fig.~\ref{figsli}), { suggests} that the 
molecular shocks arise at a stage of turbulent dissipation and not during the early phase of
the encounter, when cloud-cloud relative velocities may lead to the dissociation of SiO and CH$_3$OH.

Assuming that the internal structure of the colliding molecular clouds is highly clumpy (e.g., Falgarone 
\& Puget~\cite{falg85}), it is plausible to assume that a fraction of the kinetic energy dissipated 
during a cloud-cloud collision can cascade down to smaller scales. This would increase the {\sl turbulent} 
motions of clumps composing the end product of any cloud-cloud collision. On the simulations front, 
Kimura \&  Tosa~(\cite{kimu96}) have indeed found indications that the internal turbulence of clumpy 
colliding molecular clouds can increase after an encounter { (see also Bonnell et al.~\cite{bonn05}).}
The input kinetic energy typically involved 
in a cloud-cloud collision should be much larger in the northern spiral arm than in the nuclear ring. Once 
dissipated, this different input energies would end up producing a higher turbulence in the 
shocked molecular gas of the spiral arm ($\sigma_v\sim20-25$~km~s$^{-1}$; see Sect.~\ref{subkin}) compared to the 
nuclear ring ($\sigma_v\sim10$~km~s$^{-1}$; see Sect.~\ref{subkin}). As observed, in this scenario the largest difference 
between these two regions would be the amount of shocked molecular gas mass 
($f_{shock}^{arm}\sim$5--7$\times f_{shock}^{ring}$; see Sect.~\ref{submet}) and not the shock velocity 
regime which is here equal to $\sigma_v$ ($\sigma_v^{arm}\sim2\times\sigma_v^{ring}$; see Sect.~\ref{submet}). 

The fraction of molecular gas actually involved in the pre-starburst shocks produced by the \object{IC~342} bar 
defines the relevance of this process. We can estimate this fraction in \object{IC~342} making use of Eq.~\ref{equsio}. 
Given the similar properties of shocks in the GC and \object{IC~342} ({ Sect.~\ref{subgce}}), we can reasonably
adopt a value of X(SiO)$|_{shock}$ in \object{IC~342} similar to that estimated in the GC clouds on
$\sim$1--2~pc-scales by  H\"uttemeister et al.~(\cite{huet98}): X(SiO)$|_{shock}\gtrsim10^{-8}$. From the
values derived for X(SiO) in \object{IC~342} on { ${\sim80}$~pc} scales, we conclude that $f_{shock}\lesssim 0.1$
in the northern spiral arm and $\lesssim 0.02$ in the nuclear ring.   
In Appendix~\ref{appene}, we estimate the rate of energy dissipated by shocks in the gas over the spiral arm
region of this galaxy. While this 
estimate is rather approximate, the outcoming picture underlines the potential role of large-scale shocks at
draining energy from the gas inflowing towards galactic nuclei.   

\section{Conclusions and perspectives}
\label{seccon}

The high-resolution images showing the emission of SiO in the inner { ${r\sim320}$~pc} disk of \object{IC~342} reveal the onset of 
large-scale molecular shocks driven by the bar potential of this galaxy. The variation 
of the estimated SiO abundance inside the mapped region (from $\sim$10$^{-10}$ to $\sim$10$^{-9}$) and the
comparison with other molecular tracers (CH$_3$OH and H$^{13}$CO$^{+}$) indicate that shocks process with
uneven  { \emph{efficiency} (see Sects.~\ref{secabu} and \ref{seceff})} the molecular gas reservoir of \object{IC~342}. 
The shocks seem to arise during cloud-cloud
collisions at the stage when kinetic energy has partly dissipated in turbulent motions. The mass of
molecular gas locally involved in shocks over the spiral arm region of \object{IC~342} could amount to 10$\%$
of the dense gas reservoir. Taken together, these results underline the relevant role that large-scale
molecular shocks can play at shaping the evolution of gas disks. 

These observations illustrate the occurrence of molecular shocks in galaxies which are not related with ongoing star formation. 
SiO emission in \object{IC~342} traces a pre-starburst phase in molecular gas. Pre-starburst shocks have also
been identified to be responsible of the intense ro-vibrational H$_2$ lines detected at 2~$\mu$m in 2
prototypical mergers: \object{Arp~220} and \object{NGC~6240} (Rieke et al. \cite{riek85}). More recently, Haas
et al.~(\cite{haas05}) have reported the detection of widespread emission of the v=0--0 S(3) line of H$_2$ at
9.66~$\mu$m in the overlap region of \object{the Antennae} galaxy pair. Haas et al.~(\cite{haas05}) interpret
this emission as a tracer of shocks that will give rise to the first generation of stars in the region. In the
case of \object{the Antennae},  Haas et al.~(\cite{haas05}) 
hypothesize that shocks may not be the result of direct collisions of molecular clouds but arise instead from
an overpressured medium that remains of the collisions between H{\sc i} clouds (Jog \& Solomon~\cite{jog92}). 

In contrast to \object{the Antennae}, the bulk of the shocked gas in \object{IC~342} is not expected to produce on-site star 
formation due to the inhibiting action of strong shear over the spiral arms.  SiO emission  in \object{IC~342}
traces the sites where molecular clouds dissipate a fraction of their energy through collisions; the energy
lost helps the gas to fall to the nuclear ring where it will feed the starburst. The SiO map  of
\object{IC~342} provides a snapshot view of the pre-starburst phase during the fueling process driven by
density waves. Higher-resolution observations are required to provide new constraints on the details of how
density waves operate to produce 
molecular shocks. In particular, we expect that the efficiency of shocks changes transversally to the  
spiral arms, as cloud orbits are re-oriented by the bar potential. Because of its closeness, favourable
orientation and well defined spiral pattern, \object{IC~342} is a good target for follow-up
studies. 

High-resolution SiO imaging is key to discern the different sources of shock chemistry which are activated at
different locations and at different moments in galaxy disks during a starburst event.  Being more than a mere
tracer of {\it exotic} chemistry, SiO allows to probe unambiguously the regions where dust grains are being
destroyed in galaxies due to the action of density waves, star formation, galactic outflows  or X-rays
(Garc\'ia-Burillo et al.~\cite{buri00,buri01b}; Usero et al.~\cite{user04}). The study of the feedback
influence of these phenomena in nearby galactic disks is paramount to constrain models of evolution and
formation of galaxies at higher redshifts.

\begin{acknowledgements}
We acknowledge the IRAM staff for help provided during the observations and for data reduction. This paper has
      been partially funded by the Spanish MCyT under projects DGES/AYA2000-0927,  ESP2001-4519-PE,
      ESP2002-01693, PB1998-0684, AYA2002-01241, ESP2002-01627 and AYA2002-10113E. This research has made use
      of NASA's Astrophysics Data System and NASA/IPAC Extragalactic Database (NED). 
\end{acknowledgements}

\appendix

\section{Energy dissipation by large-scale molecular shocks in \object{IC~342}}
\label{appene}

Stellar bars can remove energy and angular momentum from the gas through gravitational torques and shocks.
Gravity torques from the stellar potential on the gas are expected to be the most relevant drivers of 
gas inflow in galaxies.  Torques created by large-scale stellar bars alone or helped in due time by other
secondary mechanisms, such as nested stellar bars, dynamical friction or viscosity, can drive gas inflow and
feed the central {\it activity} of the inner 1~kpc playground in galactic nuclei (e.g., Garc\'{\i}a-Burillo et
al.~\cite{buri05}). Large-scale shocks may also contribute significantly to the loss of energy.  In this
appendix, we make a rough estimate of the rate of energy losses produced locally by the onset of large-scale
shocks in the gas over the spiral arm region of \object{IC~342}.

A shock-front propagating in a gas medium dissipates \emph{ordered} kinetic energy into heat. For large Mach
numbers ($v_{shock}>>$~sound speed),  the kinetic energy lost per unit \emph{shocked} mass is $\sim
v^2_{shock}/2$, both in radiative and non-radiative shocks.  (e.g. Draine \& McKee~\cite{drai93}). The rate of
energy dissipated per unit \emph{total} mass due to shocks would be:       

\begin{equation}
\label{equesh}
\dot{e}_s=-\frac{f}{\tau_\mathrm{SiO}}\frac{v^2_{shock}}{2}
\end{equation}

$f$ is the total fraction of shocked gas (including the dense molecular component traced by SiO and the more 
diffuse gas), while $\tau_\mathrm{SiO}$ is the depletion timescale of SiO onto grains ($\sim10^4$~yr; Bergin et al.~\cite{berg98}).
Since gas-phase SiO depletes quickly onto grains, the shocked gas traced by SiO must have been shocked within 
a time $\tau_\mathrm{SiO}$. The expected values of $f$ range between $f_{shock}$ (if dense and diffuse gas
are assumed to be shocked to the same extent) and $\sim0.1\times f_{shock}$ (if we assume that only the dense 
gas is preferentially shocked; in this case we take a fraction of dense gas of $\sim10\%$ in \object{IC~342}
from Schulz et al.~\cite{schu01}). Assuming a typical $v_{shock}$ of 25~km~s$^{-1}$, $\dot{e}_s=
-(0.3-3)\times10^3$~km$^2$~s$^{-2}$~Myr$^{-1}$ for a corresponding range in $f=0.01-0.1$.   

Assuming quasi-circular motions, we can estimate a dissipation timescale, $\tau_s$, for the shocks to drain
the specific energy of the gas, $e$ ($e\simeq v^2_\varphi/2+\phi$, where $v_\varphi$ is the azimuthal velocity
and $\phi(r)$ is the mean gravitational potential):

\begin{equation}
\tau_s=-\frac{e}{\dot{e}_s}\times\frac{2\pi}{\theta_{arm}} 
\end{equation}

The factor $(2\pi)/(\theta_{arm})$ (where $\theta_{arm}$ is the angular width of the spiral arms at a given
radius) accounts for 
the fact that, along an orbit, shocks dissipate energy only during the time spent by the gas in the spiral 
arm. Using $^{13}$CO data, Turner \& Hurt~(\cite{turn92}) fitted a rotation curve in the inner { 320~pc} of
\object{IC~342} of the form $v_\varphi(r)=a(\sqrt{1+br}-1)$ (with $a,b$ constants); from the equation of
motion, $v^2_\varphi/r=\mathrm{d}\phi(r)/\mathrm{d}r$, we can estimate $\phi(r)$ and $e$. The value of  $e$
results to be $\simeq(2-3)\times v^2_\varphi/2$.  At the radius of GMC D { (${r\sim220}$~pc)}
$v_\varphi\sim60$~km~s$^{-1}$ and $\theta_{arm}/(2\pi)\sim0.1$, so we
obtain $e\simeq 5\times10^3$~km$^2$~s$^{-2}$. The corresponding $\tau_s$ is
$\sim16-160$~Myr, i.e. { ${\sim0.7-7}$ rotations} at the location of GMC D. 

Our estimates suggest that the energy could be drained efficiently from the gas  by large-scale shocks along
the spiral arms. While it is true that the inflow of gas is mostly constrained by the angular momentum
transfer, rather than by the energy dissipation rate (draining angular momentum is more difficult), large-scale
shocks could have indeed a non-negligible influence in the dissipation of energy of the gas on its way to the
nucleus.

\end{document}